\documentclass[twocolumn,showpacs,preprintnumbers,amsmath,amssymb,pre]{revtex4-1}

\usepackage{color}    
\usepackage{graphicx}
\usepackage{dcolumn}
\usepackage{bm}
\usepackage{subfigure}
\usepackage{amssymb}
\usepackage{multirow}
\usepackage{enumitem}
\setlist[description]{leftmargin=\parindent,labelindent=\parindent}
\usepackage[usenames,dvipsnames,svgnames,table]{xcolor}
\graphicspath{{plots/}}

\renewcommand{\vec}[1]{\mathbf{#1}}

\begin{document}


\title{Wakefields in streaming plasmas:\\
Characteristics  of the induced charge density distribution}

\author{ Zh. A.  Moldabekov$^{1, 2}$, P. Ludwig$^{1}$, J-P Joost$^{1}$, M. Bonitz$^{1}$, and T. S.  Ramazanov$^{2}$}

\affiliation{
 $^1$Institut f\"ur Theoretische Physik und Astrophysik, Christian-Albrechts-Universit\"at zu Kiel,
 Leibnizstra{\ss}e 15, 24098 Kiel, Germany}
 \affiliation{
 $^2$Institute for Experimental and Theoretical Physics, Al-Farabi Kazakh National University, 71 Al-Farabi str.,  
  050040 Almaty, Kazakhstan
  }


\begin{abstract}
Motivated by experiments on the generation of streaming plasmas in high energy density facilities, industrial setups, and fundamental dusty plasma research the plasma polarization around a test charge in streaming plasmas is considered.
The induced charge density distribution of plasma constituents is discussed for the subsonic, sonic, and supersonic regime.
Also, it is shown that the plasma polarization in the vicinity of the test charge shows different scaling in  subsonic and supersonic regimes.

\end{abstract}

\pacs{xxx}
\maketitle

\textit{Introduction:} Screening of a test charge is one of the most fundamental issues of plasma physics. 
While screening is very well understood in a equilibrium case \cite{Lampe, POP2015}, there is no consistent physical picture of this effect in streaming plasmas.
Streaming plasma attracts growing interest as typically plasma in the experiments is far from the thermodynamically equilibrium state.
Examples of plasma in such regime include complex (dusty) plasma \cite{Goree, Hanno}, dense plasma \cite {NIF, PRE2015}, so-called warm dense matter \cite{GSI}, and ultrarelativistic plasma \cite{Thoma}  generated by a beam of
energetic particles. A very active consideration of the streaming induced phenomena is in the field of complex plasmas as it allows for studying physical properties of charged many particle systems on the kinetic level \cite{Bonitz} and experiments on the generation of streaming plasmas in high energy density facilities \cite{Taccogna, Tanaka, Bastykova} as well as  industrial setups \cite{Boufendi, Boufendi2, Batryshev}. Complex (dusty)  plasma is a partially ionized plasma with an additional component of micron and submicron sized charged ``dust'' particles~\cite{Bonitz}.

For recent results on the screened charge potential in streaming plasmas see e.g. Refs. \cite{Patrick_NJP, PRE2015, CPP2015, Kompaneets, Miloch} and references therein.
A main observation is that there is a significant deviation from the Yukawa type screening.
In particular, a wake field around the test charge is formed, creating an attractive force for other like-charged particles downstream.
The topological characteristics of the wake field strongly depends on the type of the energy (velocity) distribution of the plasma particles \cite{CPP2015, Sundar} 
and on the configuration of the external electric as well as magnetic fields \cite{Joost}. 
 
An appropriate analysis of the induced charge density around the test charge gives a clear answer with respect to the correct physical picture. 
To this end, in this paper, we present the results of induced charge density around the test charge and analyse the relevant plasma polarization effects. The results are obtained from high resolution linear response calculations 
which have been validated against PIC simulations \cite{Patrick_NJP, Sundar}. 

\textit{Linear response approach:} In the experimental complex plasmas, the distribution of the flowing ions is non-Maxwellian and significantly differ from the shifted-Maxwellian \cite{Hanno2}. 
Compared to the Maxwellian (shifted-Maxwellian) case, the most prominent feature of the ion wake in the non-Maxwellian case is that the screened potential has solely a single main maximum in the trailing wake instead of an oscillatory wakefield \cite{Kompaneets}. This is the result of the ion-neutral collisions accompanied by the charge exchange effect \cite{Lampe}. In contrast, the much lighter electrons can be described by equilibrium Maxwellian distribution. 

Therefore in this work the dielectric function based on the non-Maxwellian distribution of the ions is used to calculate the induced charge density around test charge.
The dielectric function obtained in the relaxation time approximation reads \cite{Ivlev, Hanno2}

\begin{equation}\label{eq:df}
 \epsilon (\tilde {\vec k},0)=1+\frac{1}{\tilde {k}^2 \tau}+\frac{1}{\tilde {k}^2+i \nu_i M_{\rm th}\tilde {k}_z}\frac{1+\left<\xi(x)Z(\xi(x))\right>}{1+\left<\xi(0)Z(\xi(0))\right>},
 \end{equation}
where $\tau=T_e/T_n$ is the ratio of the electron temperature to the temperature  of atoms (neutrals), $\tilde {\vec k}$ is the wave vector in units of $v_{\rm th}/\omega_{\rm pi}$, $v_{\rm th}$ is the thermal velocity of atoms,
 $\omega_{\rm pi}$ is the plasma frequency of ions, $\nu_i$ is the the ion-neutral collision frequency in units of ion plasma frequency,
 the   Mach  number $M =v_d/c_s$ is  defined  as  the  ratio  of  the  ion  streaming  velocity $v_d$  and  the  ion  sound  speed $c_s=\sqrt{k_BT_e/m_i}$,
 and $\left<...\right>=\int_0^{\infty}...exp(-x)$ is the average involving following functions:

\begin{align}
 \xi(x)&=\frac{i\nu_i-M_{\rm th}\tilde {k}_zx}{\sqrt{2(\tilde {k}^2+i \nu_iM_{\rm th}\tilde {k}_z)}},\\
 Z(z)&=i\sqrt{\pi}\omega(z),\\
 \omega(z)&=e^{-z^2}{\rm Erfc}(-iz)=\frac{i}{\pi}\int_{-\infty}^{\infty}\frac{e^{-t^2}}{z-t} {\rm d}t.
 \end{align}
 
 Using dielectric function (\ref{eq:df}), the induced charge density is calculated according to the following formula:
 \begin{equation}\label{eq:den}
\tilde{n}_{\rm ind}(\tilde {\vec k})=\frac{Q_d}{e}\left(\frac{1}{ \epsilon (\tilde {\vec k},0)}-1\right),  
 \end{equation}
where $Q_d$ is the dust (or test) particle  charge.

The  computation  of  the induced charge density in  real  space  is  based  on  a  numerical  three-dimensional  Discrete  Fourier  Transformation  (3D  DFT)  
on  a  large  grid  with  resolutions  $4096 \times  4096 \times  16 384$ \cite{Book}.  In  order  to  handle  3D  grids  
of  this  size,  the  recently  introduced  high  performance  linear response program  \textsc{Kielstream} is used \cite{Book2}.

Without  loss  of generality, we consider a grain charge of $Q_d = −10^4 e$.
Specific parameters are fixed temperature ratio $\tau=100$,  and 
the  ion-neutral  collision  frequency  $\nu=0.01$, which is close to the collisionless case.
The Mach number is varied in the range $M = 0.1...2.8$. Note that for very small ion streaming velocities  the  linear  response  approach  may  not  be  applicable  due  
to  strong  dust-plasma  interactions. Also, justification of the consideration of very large values of $M$ can be problematic due to possible manifestation of the instabilities \cite{Ivlev, Kompaneets}.

\begin{figure}[h]
\includegraphics[width=0.45\textwidth]{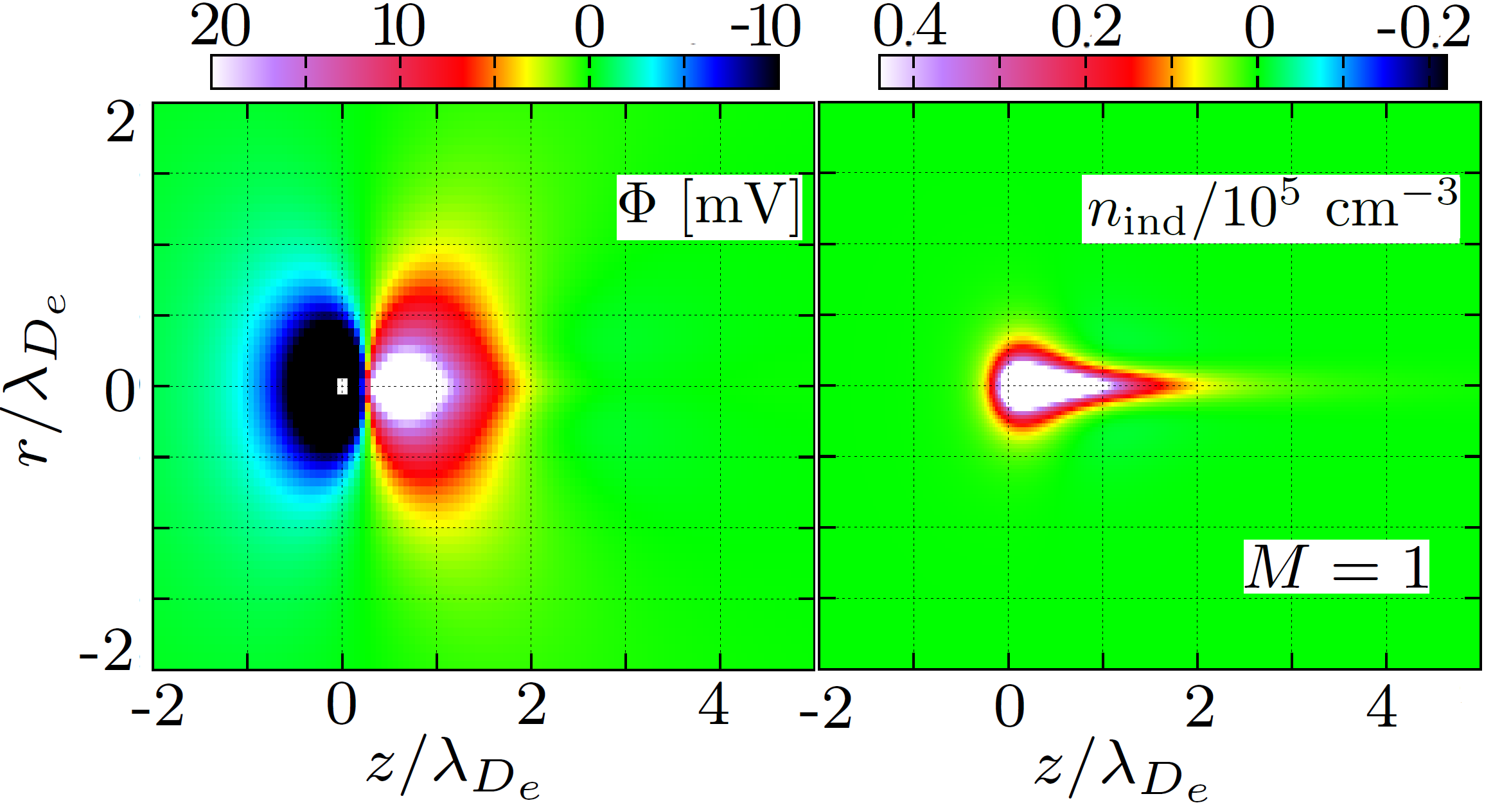}
\caption{Electric potential (left) and induced charge density (right) around the test charge at $M=1$.}
\label{fig:2}
\end{figure}

\textit{Total induced charge density distribution:} As a representative example, in Fig. ~\ref{fig:2} contour plots of the screened test charge potential and of the induced charge density distribution in the sonic case, $M=1$, are shown.
It is seen that the pattern of the induced charge density has completely different topological structure than one obtains from the intuitive picture of an ion cloud focused at some distance from the test charge.
Instead it has the shape resembling candle flame. 
 \vspace{10pt}

\begin{figure}[h]
\includegraphics[width=0.48\textwidth]{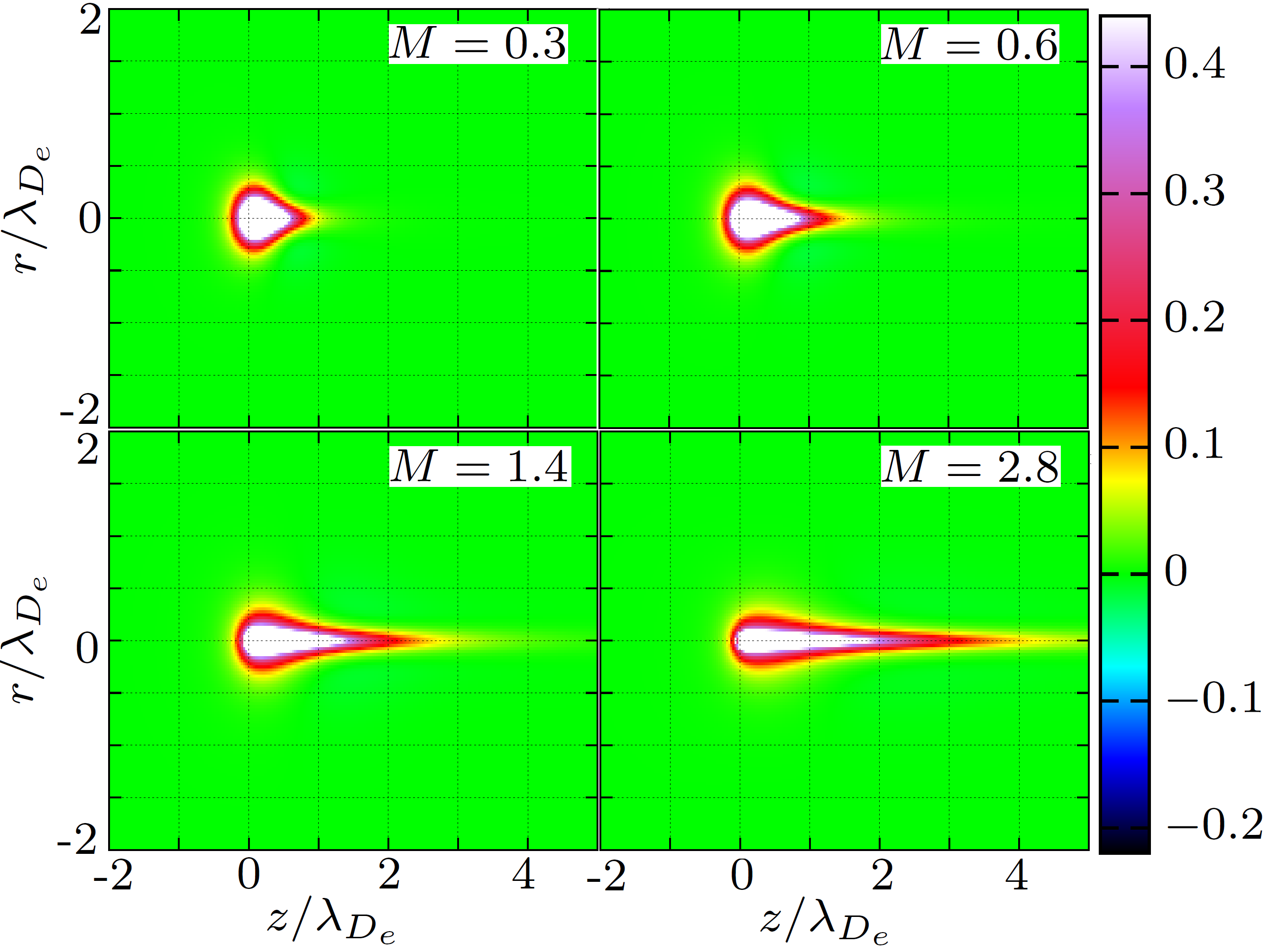}
\caption{Induced charge density in units of $10^5~{\rm cm^{-3}}$ for subsonic ($M=0.3$ and $M=0.6$) and supersonic ($M=1.4$ and $M=2.8$) regimes}
\label{fig:3}
\end{figure}

The systematic trend of induced charge density in the cases of subsonic and supersonic regimes are presented in Fig. \ref{fig:3}.
The main feature to be noticed, is that the characteristic flame type shape is preserved, but become more prolonged in the direction of the streaming with increase of the Mach number.

 \begin{figure}[h]
\includegraphics[width=0.4\textwidth]{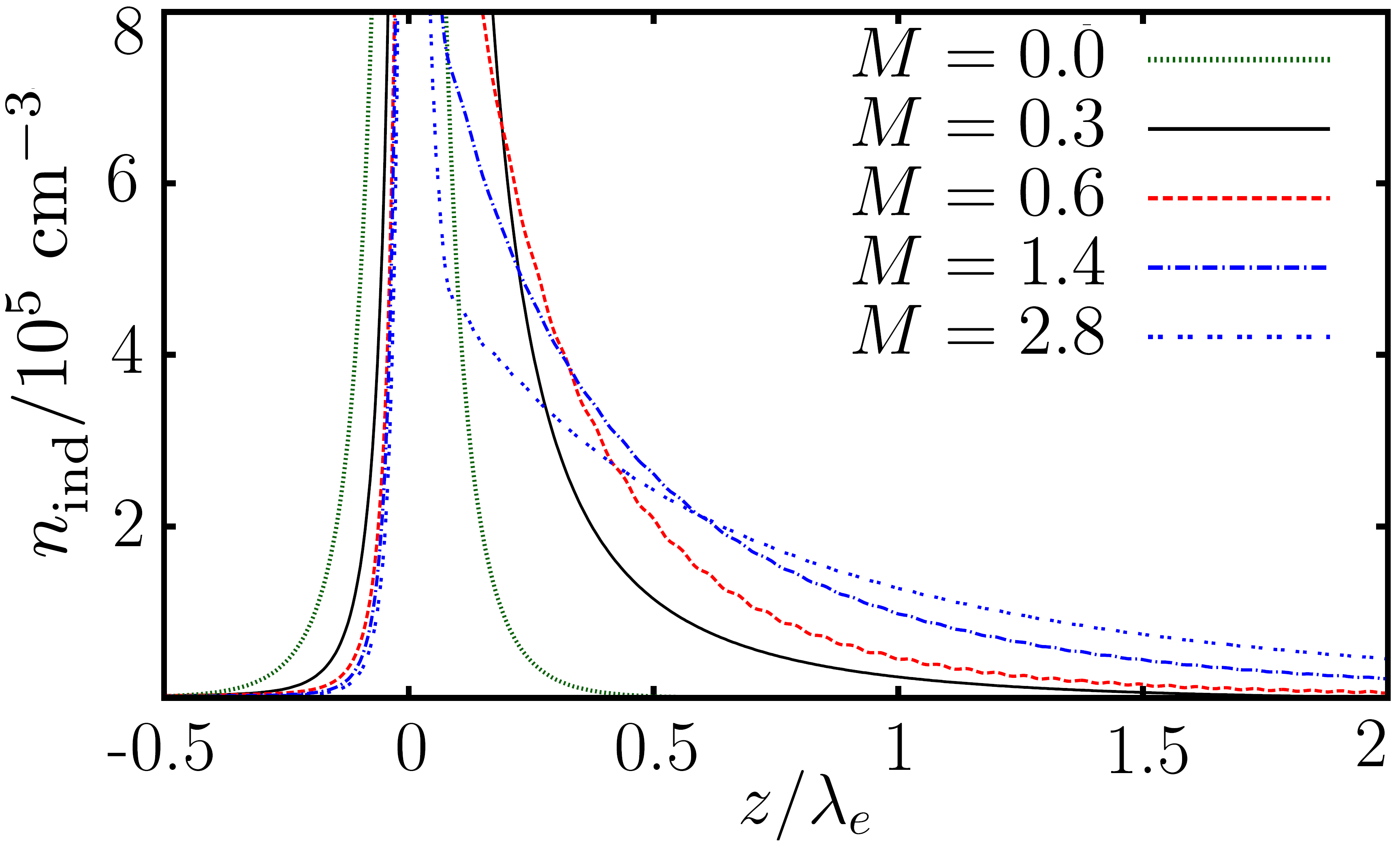}
\caption{Induced charge density around the test charge along streaming direction ($r=0$).}
\label{fig:4}
\end{figure} 

In Fig. \ref{fig:4}, the induced charge density profile along $z$ axis, which is parallel to streaming direction and passes the test particle location, is given.
With increase of the ion flow velocity $M$, in the space upstream direction of the test charge, $z<0$, the induced charge density decreases, whereas behind the test charge, $z>0$,  increases.
This results in weaker screening  at $z<0$ and appearance of an area, at $z>0$, where another like charged test particle is attracted. Most importantly, from Fig. \ref{fig:4}, one can clearly see that there is no prominent maximum (oscillatory wake) at $z>0$.

\begin{figure}[h]
\includegraphics[width=0.35\textwidth]{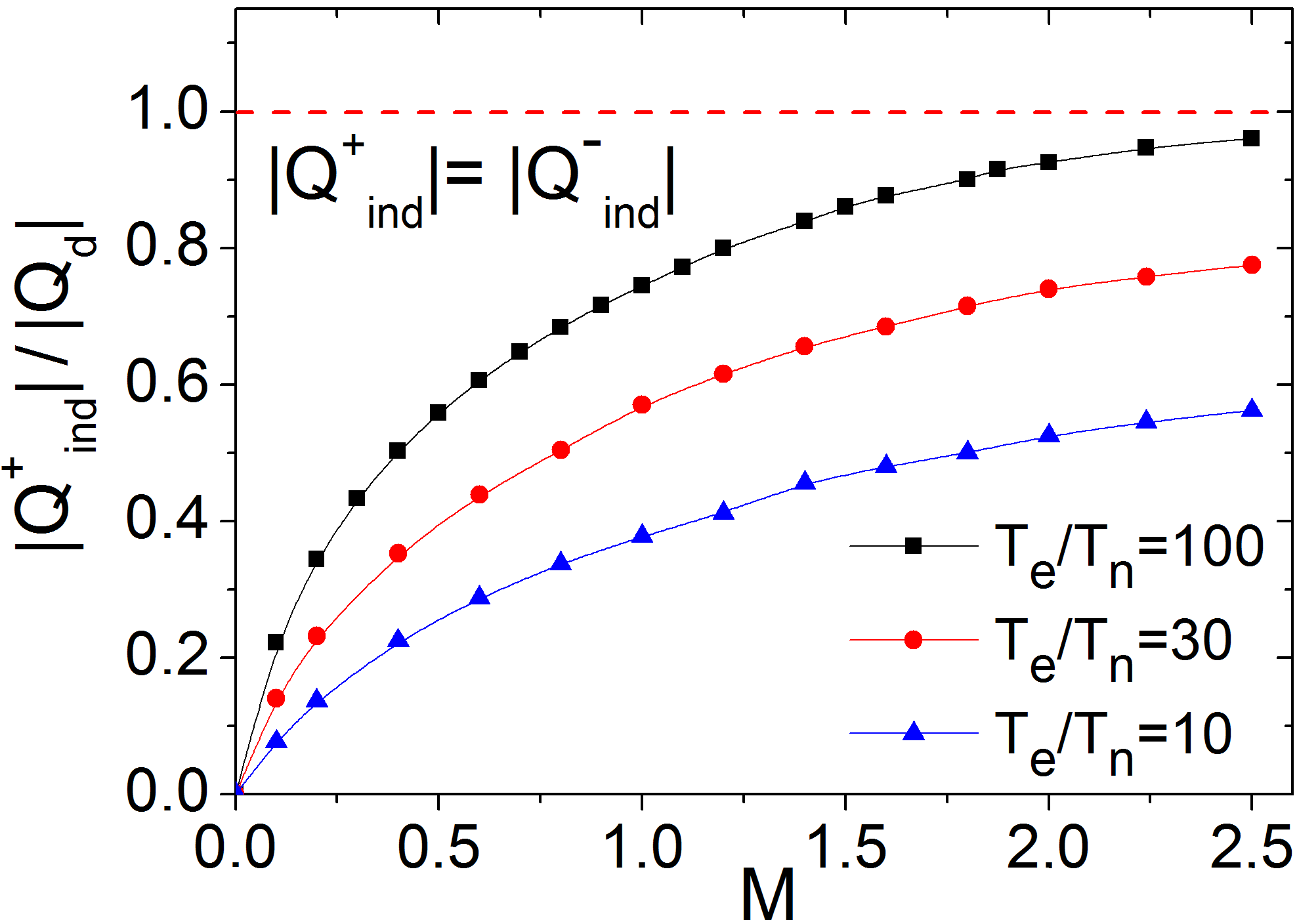}
\caption{Total value of the charge enhancement  as the function of the ion flow velocity. 
Note that absolute value of the total charge depletion around the test particle
resulted by deviation from the equilibrium (Yukawa) case is equal to the total value of the accommodated positive charge.}
\label{fig:8}
\end{figure}

In Fig.~\ref{fig:8} the value of the accommodated charge due to streaming is presented.
Large amount of the induced charge can be displaced from the vicinity of the test particle and congregated as the accommodated positive charge behind test particle, at $z>0$.
Additionally, the amount of the accommodated charge can not exceed the absolute value of the test particle charge. First, at $M<0.5$, increase in the Mach number leads to the fast increase in the total value of the accommodated charge.
After that, the total value of the accommodated charge saturates approaching $|Q_d|$.

\textit{Conclusions:} Consideration of the plasma polarization around test charge taking into account  the non-Maxwellian distribution of the flowing ions has revealed that 
the induced charge around test particle does not have neither prominent maximum nor an oscillatory pattern. 
The shape of the induced charge is ``candle flame'' type. 
The calculation of the induced charge density around test particle  has shown that the streaming leads to the polarization of the charge distribution with accommodation of 
the positive charge behind dust particle and charge depletion in direct vicinity of the test particle.

\end{document}